# Primordial planets, comets and moons foster life in the cosmos

Carl H. Gibson[a*], N. Chandra Wickramasinghe[b] and Rudolph E. Schild[c]
[a] UCSD, La Jolla, CA, 92093-0411, USA;
[b] Cardiff Univ., Cardiff, UK;
[c] Harvard, Cambridge, MA, USA

**ABSTRACT**

A key result of hydrogravitational dynamics cosmology relevant to astrobiology is the early formation of vast numbers of hot primordial-gas planets in million-solar-mass clumps as the dark matter of galaxies and the hosts of first life. Photon viscous forces in the expanding universe of the turbulent big bang prevent fragmentations of the plasma for mass scales smaller than protogalaxies. At the plasma to gas transition 300,000 years after the big bang, the $10^7$ decrease in kinematic viscosity ν explains why ~$3 \times 10^7$ planets are observed to exist per star in typical galaxies like the Milky Way, not eight or nine. Stars form by a binary accretional cascade from Earth-mass primordial planets to progressively larger masses that collect and recycle the stardust chemicals of life produced when stars overeat and explode. The astonishing complexity of molecular biology observed on Earth is possible to explain only if enormous numbers of primordial planets and their fragments have hosted the formation and wide scattering of the seeds of life virtually from the beginning of time. Geochemical and biological evidence suggests that life on Earth appears at the earliest moment it can survive, in highly evolved forms with complexity requiring a time scale in excess of the age of the galaxy. This is quite impossible within standard cold-dark-matter cosmology where planets are relatively recent, rare and cold, completely lacking mechanisms for intergalactic transport of life forms.

## 1. INTRODUCTION

Fortunately for progress in astrobiology, improved telescopes and theories reveal fatal flaws in the standard dark-energy Λ cold-dark-matter CDM hierarchical-clustering HC cosmological model for gravitational structure formation, where unwarranted assumptions about cosmic fluids predict a barren, lifeless universe. Life in the universe is virtually inconceivable, anywhere, ever, from ΛCDMHC, let alone a cosmic-scale dispersal of life seeds termed cometary-panspermia. The most important ΛCDMHC blunder is the neglect of both photon-viscosity in the plasma epoch and gas-viscosity at transition. As photons decouple from matter 300,000 years after the big bang, the phase change triggers gravitational formation of primordial planets in million-solar-mass clumps as predicted by Gibson (1996)[1] and observed independently by Schild (1996)[2]. Although it is generally accepted that the Earth is not the center of the universe, it is widely held that life originated locally and is confined to the planet Earth. According to this point of view, life is not an important cosmological process. We suggest otherwise.

Over many decades, Hoyle and Wickramasinghe (1977[3], 1982[4], 2000[5]) have pioneered the concept that cometary-panspermia explains life on Earth, citing strong spectral evidence that polycyclic-aromatic-hydrocarbons PAH are biological in origin. The close morphological similarity between terrestrial and meteoric microfossils can also be taken as proof of cosmic biological dispersal. Hydrogravitational dynamics HGD cosmology supports life as an active component of cosmic evolution from the beginning of the gas epoch soon after the big bang, Gibson, Schild & Wickramasinghe (2010)[6], Table 1.

---

[*] Corresponding author: Depts. of MAE and SIO, CASS, *cgibson@ucsd.edu*, **http://sdcc3.ucsd.edu/~ir118**

Table 1. HGD structures formed in the plasma and gas epochs as a function of time after the big bang event. Because the universe is expanding, gravitational instability favors density minima. Structure formation is by fragmentation as the minima form voids.

| Protosuperclusters and voids | 30,000 years ($10^{12}$ s), plasma |
|---|---|
| Protogalaxies on vortex lines | 300,000 years ($10^{13}$ s), plasma |
| Planets in clumps, stars, leading to life | 330,000 years ($10^{13}$ s), gas |
| Intelligent life | ~13.7 Gyr ($4 \times 10^{17}$ s) |

Cometary-panspermia CP and hydrogravitational dynamics HGD cosmology are complimentary concepts that must be merged if one is to explain the flood of observations coming in from a host of new space telescopes and greatly improved ground based telescopes. Each of these concepts is needed to fully explain the other. Without HGD, CP has no life seeds to disperse and no means to disperse them beyond a few planets near a star if, against all odds, life should appear on one of these few planets. Without an abundance of primordial life forms, biological PAH and complex organics, HGD offers no explanation for numerous observations too obvious to be ignored; for example, most blatantly, the existence of life, and intelligent life, on Earth.

## 2. THEORY

Gravitational instability is absolute. Suppose a large volume of fluid exists at rest with uniform density $\rho$. All the fluid within radius ct will begin to move toward the origin $\mathbf{x} = 0$ for time $t > t_0$ if a non-acoustic mass perturbation of size $M' = +\delta\rho L^3$ is introduced at $t = t_0$, and all the fluid will begin to move away from the origin if $M' = -\delta\rho L^3$, where c is the speed of light and L is a radius smaller than the Jeans acoustic scale $L_J$ but larger than the largest Schwarz scale $L_{SXmax}$. It can be shown (Gibson 2000)[7] that an isolated mass perturbation grows or decreases exponentially with time squared according to the expression

$$M'(t) = |M'(t_0)| \exp [\pm 2\pi\rho G t^2]$$

where the gravitational (free fall) time scale $\tau_g = (\rho G)^{-1/2}$. Nothing much happens at the origin before the free fall time $\tau_g$ arrives. For $M'$ positive a sudden increase in density occurs near the "cannonball" origin as mass piles up. Hydrostatic equilibrium is established as local pressure increases to balance gravity. Density and temperature increase and hydrodynamic effects, reflected by the Schwarz scales, become relevant. For $M'$ negative, density near the origin suddenly decreases and a void appears surrounding the "vacuum beachball" origin at time $t \sim \tau_g$. The void then propagates radially outward as a rarefication wave limited in speed by the speed of sound. Matter near the void falls radially away due to gravity. Thus, in an expanding universe, gravitational structure formation occurs mostly by fragmentation since the expansion of space favors void formation and resists condensation. Formation of voids in the plasma epoch causes the prominent sonic peak in the cosmic microwave background CMB temperature anisotropy spectrum, not acoustic oscillations of plasma trapped in gravitational potential wells of condensed CDM seeds.

The viscous Schwarz scale $L_{SV} = (\gamma \nu / \rho G)^{1/2}$ is the length scale at which viscous forces match gravitational forces, where $\gamma$ is the rate-of-strain, $\nu$ is the kinematic viscosity of the fluid and G is Newton's gravitational constant. The turbulent Schwarz scale $L_{ST} = \varepsilon^{1/2}/(\rho G)^{3/4}$, where $\varepsilon$ is the viscous dissipation rate of turbulent kinetic energy. The diffusive Schwarz scale $L_{SD} = (D^2/\rho G)^{1/4}$, where D is the diffusivity of the fluid.

The Jeans acoustic scale $L_J = V_S \tau_g$. Only the Jeans (1902) criterion that gravitational structure formation cannot occur for scales $L < L_J$ is used in the standard cosmological model. This is its fatal flaw. Jeans neglected all effects of viscosity, turbulence and diffusivity, and so does ΛCDMHC cosmology. Because the scale of causal connection $L_H = ct$ is smaller than the Jeans scale during the plasma epoch, no structure can form. Thus it became necessary to invent cold dark matter CDM, which is a mythical substance with the property that its Jeans scale $L_J$ is smaller than $L_H$ because it is cold. The standard model assumes CDM is nearly collisionless, like neutrinos, which means its diffusivity D is very large so that its $L_{SD}$ scale during the plasma epoch exceeds $L_H$. Thus, whatever CDM is, it cannot condense during the plasma epoch according to the hydrogravitational dynamics HGD Schwarz scale criteria (Gibson 2009a, b, 2010)[8,9,10].

## 3. STRUCTURE FORMATION IN THE EARLY UNIVERSE

Figure 1 illustrates schematically the differences between HGD cosmology and ΛCDMHC cosmology during the plasma epoch, soon after mass became the dominant cosmological component at time $\sim 10^{11}$ seconds after the big bang event over energy. From HGD, 97% of the mass at that time is non-baryonic, with the weakly collisional properties and mass of neutrinos (green). The rest (yellow) is hydrogen-helium plasma (protons, alpha particles and electrons). The total mass is slightly less than required for the universe to be flat, leading to an eventual big crunch. For ΛCDMHC the anti-gravity dark-energy density eventually dominates and accelerates the expansion of an open universe as the kinetic-energy and gravitational-potential energy densities approach zero.

In Fig. 1, CDM seeds (top left) cannot possibly condense because this non-baryonic dark matter NBDM (green) is nearly collisionless and, therefore, super-diffusive. Neither can NBDM hierarchically cluster HC to form potential wells, into which the plasma (yellow) can fall and produce loud acoustic oscillations (top right). Gravitational structure formation begins in HGD (bottom left) as fragmentation at superclustermass scales ($\sim 10^{46}$ kg) to form empty protosuperclustervoids, shown as open circles. Because the universe is expanding, structure formation is triggered at density minima to form voids when viscous forces permit it. Voids expand as rarefaction waves limited by the speed of sound, which in the plasma epoch is $c/3^{1/2}$, where c is the speed of light. Fragmentation continues at smaller and smaller mass scales to that of protogalaxies ($\sim 10^{43}$ kg). Weak turbulence develops at the expanding void boundaries. This weak plasma turbulence determines the morphology of plasma protogalaxies, which then fragment along turbulent vortex lines because the rate of strain $\gamma$ is maximum on vortex lines. Consequently, the plasma-turbulence Nomura scale $\sim 10^{20}$ m determines protogalaxy sizes at transition to gas[11].

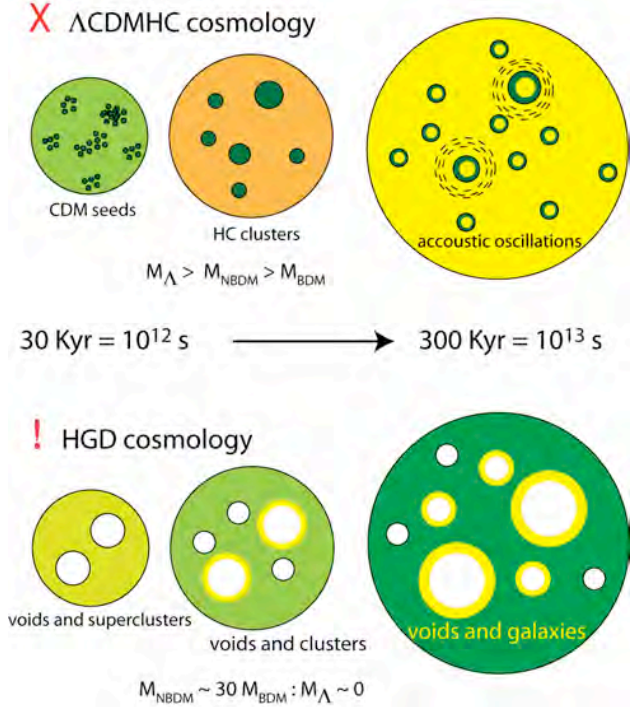

Figure 1. Comparison of plasma epoch behavior of the universe for the incorrect **X** standard ΛCDMHC model (top) versus the correct **!** hydrogravitational dynamics HGD model (bottom). Condensation of the nearly collisionless non-baryonic dark matter to form CDM seeds that cluster (top) is physically impossible. Structure formation occurs by fragmentation (bottom) when the increasing scale of causal connection $L_H = ct$ matches the viscous Schwarz scale $L_{SV}$. This occurs at $t_0 = 10^{12}$ s (30 Kyr). Proto-galaxies produced at plasma-gas transition[11] have Nomura turbulence morphology and length scales reflecting $t_0$.

Figure 2 illustrates schematically the two cosmologies in the gas epoch from 300 Kyr to 300 Myr ($10^{13}$ -$10^{16}$ s), which is often termed the dark ages for ΛCDMHC (top) because this is the time required for the first star and the first planets to appear in this cosmology. The temperature of space has fallen to a few °K that will freeze any gas. The density has decreased by a billion. Any life arising on the handful of planets produced as stars form from gas would likely be blasted out of existence by superstars so powerful they re-ionize all the plasma of the universe as they explode. Extra-terrestrial (and terrestrial) life is virtually impossible by the standard ΛCDMHC model. Life would be extremely rare and confined to local star systems.

For HGD cosmology (Fig. 2 bottom), the $10^{13}$ -$10^{16}$ s interval has many stars and warm planets, and is the optimum time period for life to appear and for its first seeds to be widely scattered on cosmic scales. Plasma-protogalaxies form near the end of the plasma epoch by fragmentation along turbulent vortex lines. Gas-protogalaxies fragment into $10^{36}$ kg Jeans mass clumps of $10^{24}$ kg planetary-mass clouds that shrink and eventually freeze solid as H-$^4$He gas planets (termed primordial-fog-particles PFPs, or micro-brown-dwarfs μBDs) as they cool.

This is the first gravitational condensation. The planet-clump density $\rho_0 = 4 \times 10^{-17}$ kg m$^{-3}$ is that of the plasma when it first fragmented at $10^{12}$ s. It is no coincidence that this is the observed density of old globular star clusters OGCs. Clumps of planet-mass clouds are termed proto-globular-star-clusters PGCs. No dark age exists for HGD because the first stars appear in a gravitational free fall time $\sim(\rho_0 G)^{-1/2} = 2 \times 10^{13}$ s or less at temperatures too high for it to be dark anywhere. The process of star formation is binary-planet-mergers within these PGC clumps of primordial-fog-particle PFP planets.

In HGD cosmology all stars form within PGC clumps of planets. The empty volume without planets is termed the Oort cavity, with size $(M/\rho_0)^{1/3}$, where M is the mass of the central star or stars produced by planet merging from a mass density of planets $\rho_0$. This is $3\times10^{15}$ m for a solar mass, which closely matches the observed distance to the Oort cloud of long period comets. From HGD, planets continue to fall into the Oort cavity from its inner boundary, eventually overfeeding the central star or stars to the point that they explode as supernovae. Supernovae are the source of carbon, oxygen, nitrogen etc. that are needed to produce life. We see that all the conditions for the first life in the big bang universe are achieved soon after the plasma to gas transition. Because many of the planets and their moons and meteorite fragments will be relatively warm, many domains of liquid water will exist in this early period as primordial soup kitchens. Once the first templates for life appear, they will rapidly be transmitted to other planets and their associated venues so that life can evolve by mutation and evolution as we see it on Earth.

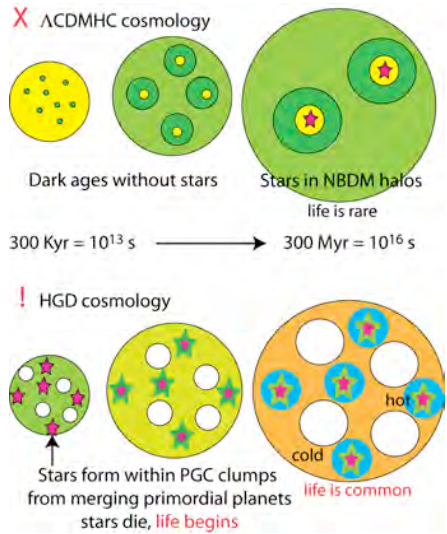

Figure 2. Gas epoch for 300 Kyr to 300 Myr, corresponding to the dark-ages period for ΛCDMHC (top) before the first stars and planets. Astrobiology would be very different in ΛCDMHC because, even if life were to form, it would be confined to a handful of planets formed simultaneously and close to their stellar mother, with no common cosmic ancestors. Life in HGD cosmology (bottom) begins immediately with excellent cosmical mixing within PGC clumps at all stages, some mixing between clumps and within galaxies, and progressively less mixing at larger scales. The non-baryonic dark matter (probably neutrinos) fragments at its LSD scale after the plasma epoch to form galaxy cluster halos (green bottom center). Hot x-ray halos with small mass surround the clusters (blue bottom right). Radio telescope detection (Rudick et al. 2008)[9] of completely empty supervoids with scales ~ $10^{25}$ m contradict the ΛCDMHC model (top), which makes small voids that are not empty at its last stage rather than its first.

Figure 3 shows optical band images from the Hubble Space Telescope HST of the Helix planetary nebula PNe, one of the closest PNe to the Earth at a distance of ~$6\times10^{18}$ m[12]. The central white dwarf is constantly fed PFP planets as comets falling in from the Oort cavity boundary, which it converts to carbon. Part of the gas of the comet-planets released to the star is expelled as a plasma beam by the rapidly spinning white dwarf. The plasma beam partially evaporates and photoionizes the nearest surrounding planets to ~ $10^{13}$ m scales (depending on the mass of the frozen planet) so they are detectable. When the white dwarf mass exceeds the Chandrasekhar limit of 1.44 solar mass ($2.2 \times 10^{30}$ kg) it explodes with large predictable brightness, making it an excellent standard candle for distance estimates.

If the line of sight to the Earth intersects an evaporated planet atmosphere it is slightly dimmed, otherwise it is not dimmed (red circles) and no dark energy is needed to explain the dimming. Thus dark energy $\Lambda$ is a systematic SNeIa dimming error of dying carbon stars embedded in clumps of primordial planets, not 70% of the density of the present universe as claimed by $\Lambda$CDMHC cosmology. A similar systematic error in the age of the universe (Sandage et al. 2006)[9,10] can also be explained as systematic SNeIa dimming errors due to PFP planet atmospheres (bottom left, red circles). Rather than 16 Gyr, the corrected age is 13.7 Gyr, consistent with the value estimated from cosmic microwave background evidence.

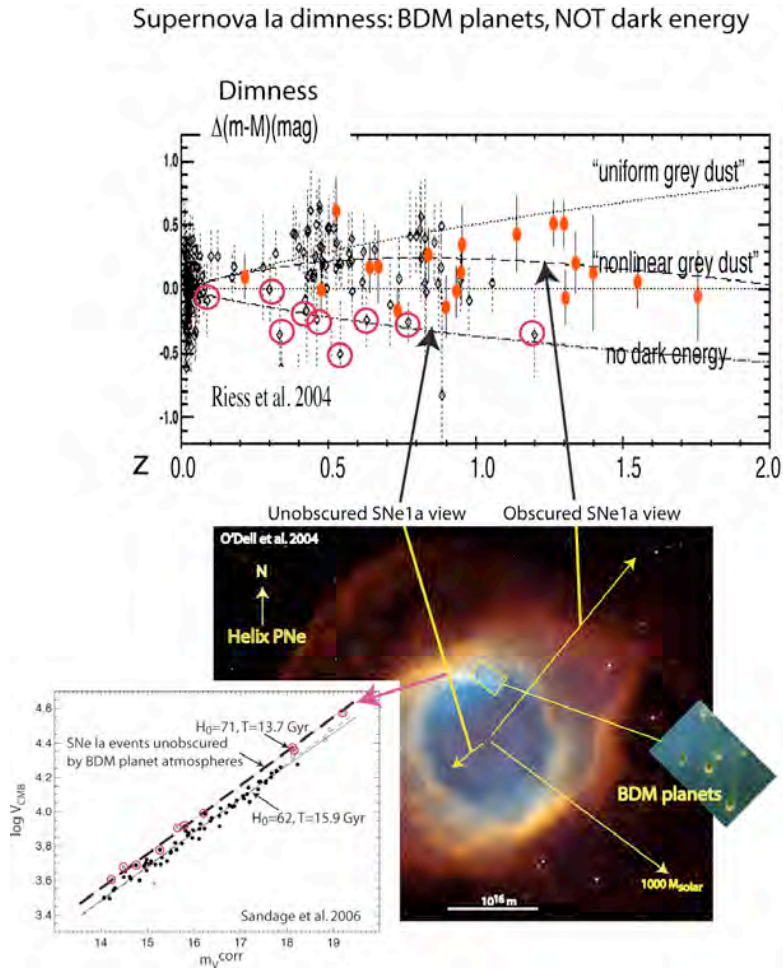

Figure 3. Helix Planetary Nebula (O'Dell et al. 2004) interpreted using HGD cosmology (Gibson 2010, Fig. 8)[10]. At the top is supernova Ia (SNeIa) dimness evidence of dark energy (red ovals) not taking effects of dark matter planets in clumps as the source of all stars and the dark matter of galaxies. Unobscured SNeIa lines of sight (red circles) are brighter, suggesting $\Lambda$ is a systematic dimming error corrected by HGD cosmology. A similar systematic dimming error due to evaporated PFP atmospheres is suggested at bottom right for the age of the universe (Sandage et al. 2006)[9,10] inferred from the dimmed SNeIa to be 16 Gyr: corrected (red circles) to 13.7 Gyr. A yellow arrow (bottom right) shows the radius of a sphere containing a thousand solar masses of PFP planets.

Most of the evaporated planets of Helix in Fig. 3 are obscured by polycyclic aromatic hydrocarbon PAH dust. Only 5000 optical planets can be seen. Infrared space telescopes such as Spitzer reveal 40,000 or more, as well as planet protocomets falling into the central star.

## 4. EVIDENCE OF PRIMORDIAL PLANETS FROM INFRARED TELESCOPES

Figure 4 shows the Helix PNe in both optical (bottom) and infrared (top) frequencies (Matsuura et al. 2009)[13]. Numerous planets and all of the planet-protocomets are quite invisible in the HST optical frequencies.

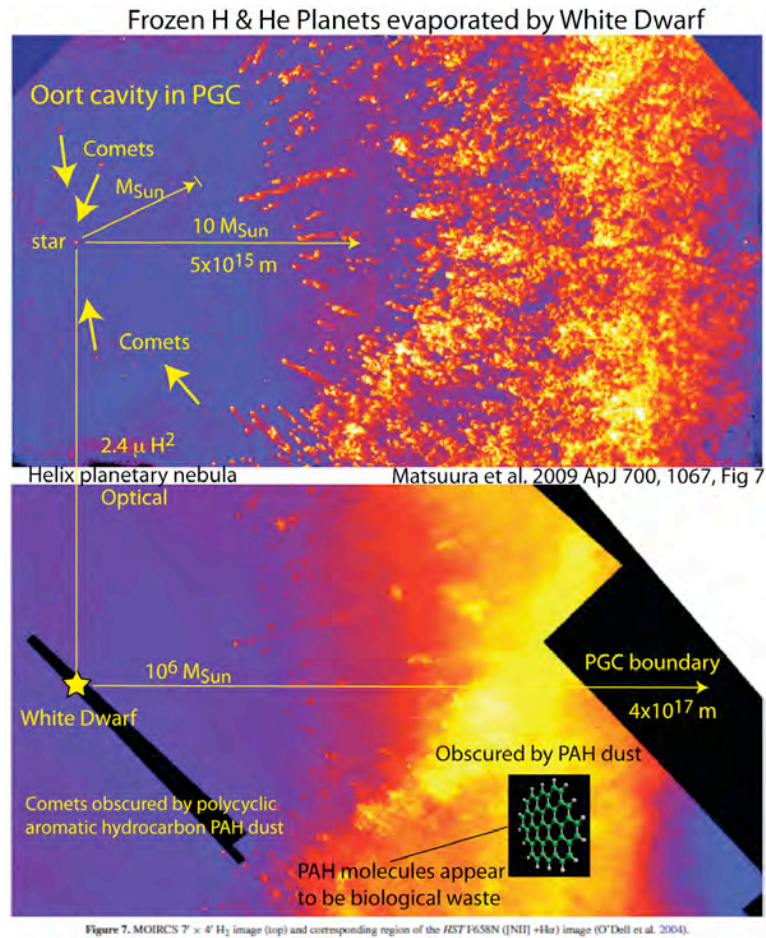

Figure 4. Helix PNe reveals partially evaporated dark matter planets (top) at the hydrogen molecule $H^2$ wavelength 2.4 μ (Matsuura et al. 2009)[13] that are obscured to an extent that increases with radius from the central white dwarf star, suggesting PAH dust has been evaporated from the planets as well as their originally frozen H-He gas. A typical PAH molecule is shown (bottom right), which has the structure of carbohydrate oils used as food by terrestrial organisms.

Note from Fig. 4 that planets at the Oort cavity boundary are sorted according to their amounts of biological activity by the radiation pressure, which increases with PAH content because the heat transfer increases with the size of the atmosphere.

Figure 5 is a similar Helix PNe comparison (Meaburn & Boumis 2010)[14], showing clumps (globulettes)[15] of PFPs revealed in the infrared that are invisible in the optical. Clumping is easy to understand from HGD since the plasma beam radiating from the white dwarf magnetic pole is responsible for the partial evaporation and photoionization of the dark matter frozen gas planets it intersects. A clockwise helical track of evaporation can be traced in Helix, which accounts for its name. The large atmospheres of the irradiated planets causes frictional forces between planets whose atmospheres overlap that may cause them to clump and in some cases merge.

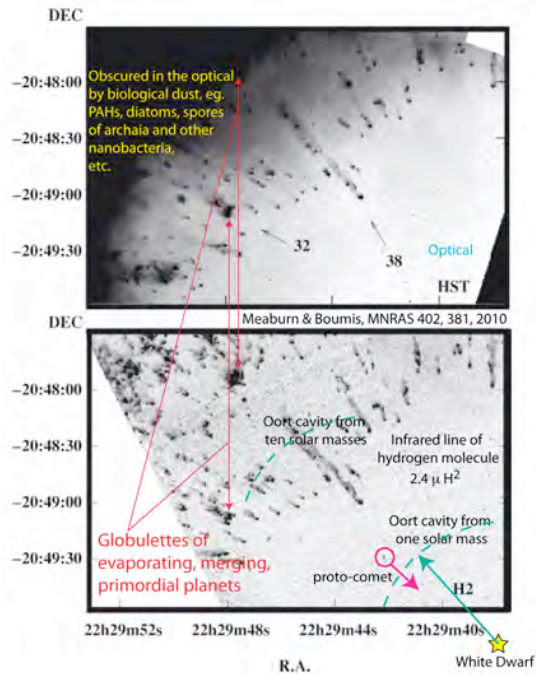

Figure 5. Helix PNe at optical frequencies (top) compared to the infrared molecular hydrogen band (bottom). Globulettes[15] of evaporating, merging primordial planets are inferred from vertical red arrows left of center connecting optical and infrared objects.

Earth is constantly bombarded by meteors. Carbonaceous meteorites such as Murchison contain ample evidence of ancient extra-terrestrial life. As shown in Figure 5, the morphology of fossils contained is that of organisms that were once alive, but most species are unknown or different from those on this planet. The complexity of the organic compounds detected in a sample of the 100 kg object is estimated to exceed a million[16], a significantly larger chemodiversity of its source compared to that of Earth. Evidence suggests Murchison is older than the Earth. Images of cyanobacteria from Muchison and Orgueil meteorites are available at Richard Hoover's website, along with numerous references, http://www.panspermia.org/hoover4.htm.

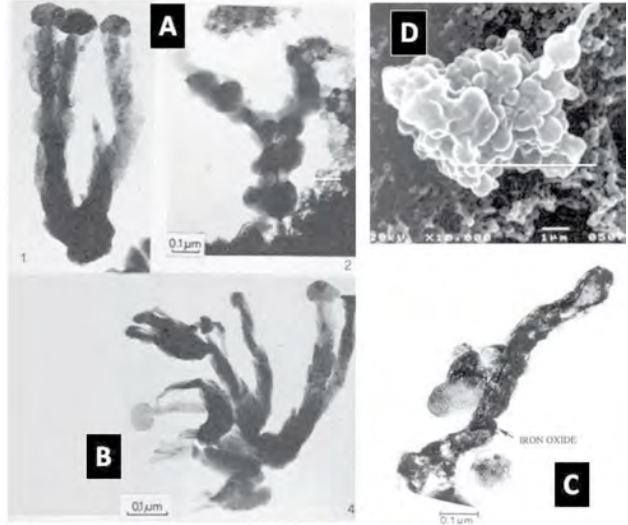

Figure 6. Microbial fossils from comets. Images A, B are bacterial microfossils in the Murchison meteorite (courtesy Hans Pflug), C is a bacterial fossil within a Brownlee particle and D is a clump of cocci and a bacillus from dust collected using a cryoprobe from 41km in the atmosphere. References and credits are given in Wickramasinghe et al., 2010[17].

Figure 7ab shows a much more mysterious organism, taken to be extraterrestrial and potentially an important link to the origin of life. It is the Red Rain material that fell on Kerala, India, in 2001, soon after a sonic boom announced the arrival of a meteor[18].

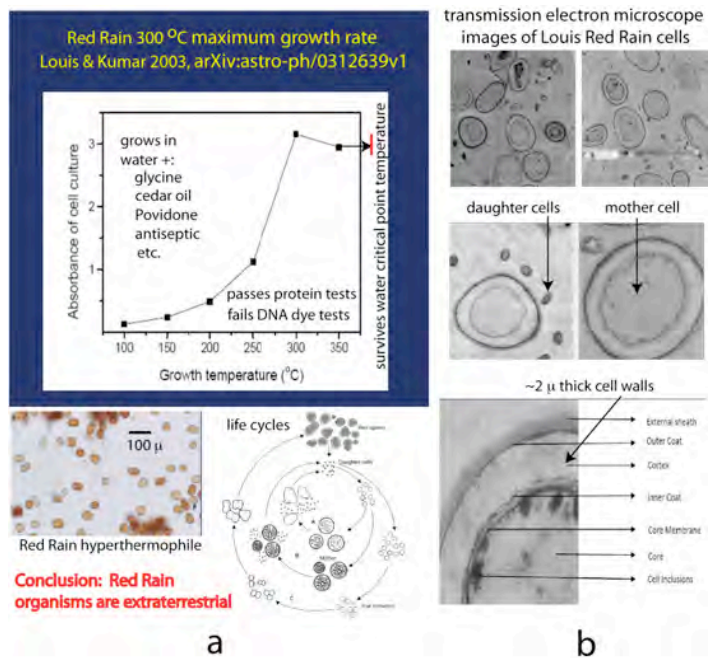

Figure 7ab. a. Louis and Kumar study[18] of Kerala Red Rain organisms. Maximum growth rates occur at a record 300°C, far above survival temperatures of known terrestrial hyperthermophiles. Survival temperatures exceed the water critical point 374 °C (red line), which is suggested[19] as optimum for the formation of life chemicals. b. Transmission electron microscope images of Louis Red Rain cells show complex inner structures, mother and daughter cells, and highly variable cell wall thicknesses at various life cycle stages[20].

The Louis and Kumar (2003)[18] investigations, supported by independent studies of Gangappa et al.[20], justify a conclusion that Red Rain organisms are living extraterrestrial hyperthermophiles of unknown provenance. We consider it likely that these organisms are of extraterrestrial origin delivered to the Earth by the mechanism of cometary panspermia. Complex life cycles (Fig. 7a bottom right, TEM images Fig. 7b) and protein tests imply DNA regulation. However, the organism fails L-amino acid based DNA dye tests, suggesting Red Rain organisms may be representative of a "shadow biosphere" of opposite chirality to normal terrestrial life. The handedness of sugars and amino acids are arbitrary, so DNA life should appear in either R or D forms. Figure 8 shows further results from other laboratories that replicate and extend the Louis and Kumar[18] results.

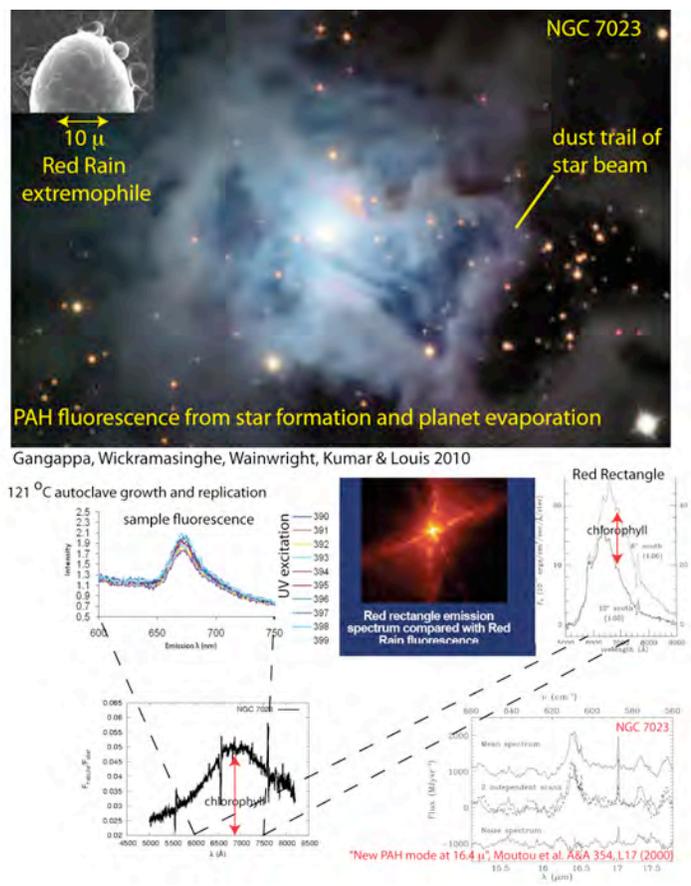

Figure 8. Polycyclic aromatic hydrocarbon PAH evidence of Red Rain organisms in the reflection nebula NGC 7023, the Red Rectangle, and in laboratory replications of the Louis growth results to 121 °C. A new PAH mode in the characteristic Red Rain frequency band is reported by Moutou et al. (2000)[21]. Chlorophyll lines appear in both 7023 and the Red Rectangle.

Although a cosmic origin of the Kerala red rain cells is still regarded as controversial there is little doubt that their properties make them uniquely suited to survival (after first conception and evolution) in the hot, high pressure interiors of primodial planets of the type we have discussed. Moreover, it is entirely possible that a subset (perhaps half) of the primordial planets in the universe harbor life of an opposite chirality to ours – a disjoint "shadow" biosphere. Injections via cometary panspermia of life of the opposite chirality may as a rule go unnoticed because of their lack of interaction with terrestrial life. Conspicuous events such as the red rain may be evidence of rare instances when this type of life form manifests itself. In this context it should be noted that reports of red rain have recurred throughout history going back to biblical times; for example, in numerous accounts of rains of blood (McCafferty[23]). The apparent localization of the episode to the small area in the state of Kerala India could be due in part to an observational selection effect. The state of Kerala is unique in India with a nearly 100% literacy rate. It is therefore no surprise that the citizens of Kerala recorded an event that may have gone unnoticed in the rest of southern India. Kerala as a tropical Indian Ocean peninsula is also particularly subject to meteor dust nucleation of strong rains during the monsoon season.

## 5. DISCUSSION

Table 1 and Figures 1 and 2 review the large differences between cosmologies with (HGD) and without (ΛCDMHC) fluid mechanics. Figures 3, 4 and 5 of the Helix planetary nebula support the HGD conclusion that the dark matter of galaxies is primordial planets in protoglobularstarcluster PGC clumps of a trillion. Figure 6 from the Murchison carbonaceous meteorite shows some of the first strong evidence of cometary panspermia at cosmic scales of extraterrestrial microfossils, as championed for many decades by N. C. Wickramasinge[17] and F. Hoyle[3,4,5].

Figures 7ab and 8 describe Red Rain evidence that extraterrestrial life exists with resistance to temperatures and chemicals terrestrial life cannot tolerate, at temperatures exceeding the critical point of water. The critical conditions of water are to astrobiology as the Planck conditions are to the big bang; that is, ideal. Chirality of DNA suggests a binary biosphere on PGC scales.

Figure 9 suggests a likely scenario for first life to occur in the hot gas primordial planets of HGD cosmology. The universe cools as it expands and the primordial planets will also cool by radiative heat transfer to space. Stars and supernovae at protogalaxy cores supply stardust chemicals of life to primordial planet clouds as they cool toward the critical temperature of water 674 K.

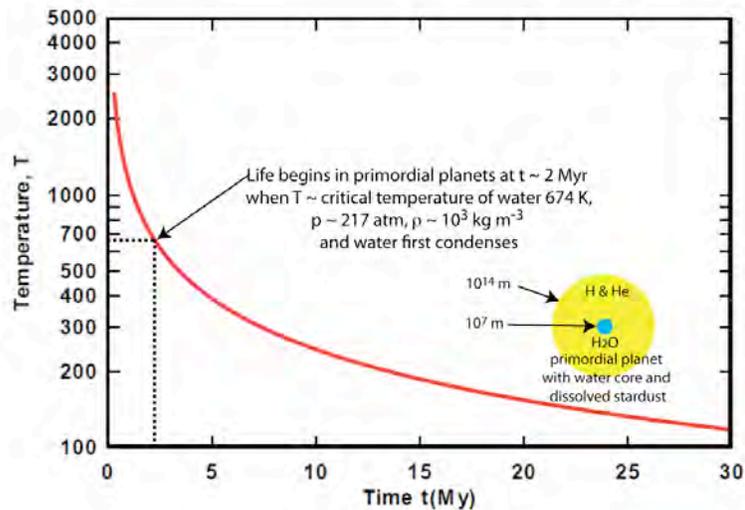

Figure 9. The optimum time and conditions for first life to form in primordial planets is when water first condenses in the presence of the necessary stardust chemicals C, O, N, P etc. following the first stars and supernovae. This occurs at approximately 2 million years after the big bang event[22]. Because of the strongly reducing hot hydrogen atmosphere (yellow) of the initial PFPs, iron and nickel particles could nucleate a rain of critical state water to form the planet core (blue).

Metallic iron and nickel from SNeII events may nucleate critical water rain that collects in an ocean at planet cores (blue in fig 9). Each rain drop is a site for critical point carbohydrate formation. Iron and nickel cores are repeatedly recycled and melted from planet mergers as they grow larger and form stars. Stratified layers of liquids and gases form above the primordial planet core, including water. From Fig. 9, critical temperature water should start to condense and form life chemicals, and life, at the core of the planets (with dissolved life chemicals C, O, N, P, Si, etc.) at about 2 million years.

### 6. CONCLUSIONS

Theory and observational evidence show that the standard dark-energy cold-dark-matter hierarchical-clustering ΛCDMHC model for cosmology must be replaced by hydrogravitational dynamics cosmology HGD. An important result of HGD cosmology is its prediction that the dark matter of galaxies is primordial planets PFPs in clumps PGC. All stars form in these PGC clumps as PFPs merge. The optimum time for first life and the spreading of the seeds of life is very early while the planets are warm and the universe is dense.

We suggest the time of first life was at ~ 2 Myr when the universe cooled sufficiently for water to condense, with stardust fertilizer C, N, O, P etc., at the core of the hydrogen helium primordial planet clouds, Fig. 9. The critical temperature of water 674 K approximately matches the breakdown temperature of amino acids needed for DNA, and permits the high speed chemical kinetics necessary for living organisms and their complex chemicals to rapidly develop and evolve the highly efficient DNA mechanisms we see on Earth. Chlorophyll catalysts for converting carbon to food can resist such high temperatures. A shadow biosphere with reversed DNA chirality is suggested by studies of the Red Rain organism, whose DNA is undetected by standard methods but whose life cycles and astrophysical signatures imply DNA capabilities.

Part of the astrobiological processes that produce life in primordial planets is the distribution of the templates or seeds of life by the plasma jets of the stars the planets produce, as well as the jets from active galactic nuclei that devour millions of stars and eject to other galaxies trillions of life infested planets and planet clumps. The biosphere and shadow biosphere therefore extend to all material produced by the big bang, about $10^{80}$ planets. Biological processes are extremely efficient at converting the carbon of planets to organic chemicals and their fossils, as we see on Earth. With high probability, life did not begin on Earth, or on any of the $10^{18}$ planets of the Galaxy, but was more likely brought to Earth and the Milky Way by cometary panspermia. Biology and medicine are thus subsets of astrobiology on cosmic scales yet to be determined by future studies.

### REFERENCES

1. Gibson, C.H., "Turbulence in the ocean, atmosphere, galaxy and universe," Appl. Mech. Rev. 49, no. 5, 299–315, 1996.
2. Schild, R., "Microlensing variability of the gravitationally lensed quasar Q0957+561 A,B," *ApJ* **464**, 125, 1996.
3. Hoyle, F. and Wickramasinghe, N.C., *Nature* **270**, 323, 1977.
4. Hoyle, F. and Wickramasinghe, N.C., "Proofs that Life is Cosmic," *Mem. Inst. Fund. Studies Sri Lanka* **1** (www.panspermia.org/proofslifeiscosmic.pdf), 1982.
5. Hoyle, F. and Wickramasinghe, N.C. *Astronomical Origins of Life: Steps towards Panspermia*, Kluwer Academic Press, 2000.

6. Gibson, C. H., Schild, R. E. & Wickramasinghe, N. C., "The origin of life in primordial planets," International Journal of Astrobiology, in press, 2010, arXiv:1004.0504.
7. Gibson, C.H., "Turbulent mixing, diffusion and gravity in the formation of cosmological structures: The fluid mechanics of dark matter," *J. Fluids Eng.* **122**, 830–835, 2000.
8. Gibson, C. H., *New Cosmology: cosmology modified by modern fluid mechanics*, Amazon.com, Inc., ISBN 978-1449507657, 2009a.
9. Gibson, C. H., *New Cosmology II: cosmology modified by modern fluid mechanics*, Amazon.com, Inc., ISBN 978-1449523305, 2009b.
10. Gibson, C. H., "Turbulence and turbulent mixing in natural fluids," *Physica Scripta Topical Issue, Turbulent Mixing and Beyond 2009*, in press 2010, arXiv:1005.2772v4.
11. Gibson, C. H., "Cold Dark Matter Cosmology Conflicts with Fluid Mechanics and Observations," *J. Appl. Fluid Mech.* **1(2)**, 1-8, 2008, arXiv:astro-ph/0606073.
12. Gibson, C.H. & Schild, R.E, "Interpretation of the Helix Planetary Nebula using Hydro-Gravitational-Dynamics: Planets and Dark Energy," 2007, arXiv:astro-ph/0701474.
13. Matsuura M. et al., "A 'Firework' of H2 knots in the planetary nebula NGC 7393 (The Helix Nebula)," *ApJ* **700**, 1067-1077, 2009.
14. Meaburn, J. & Boumis, P., "Flows along cometary tails in the Helix planetary nebula NGC 7293," Mon. Not. R. Astron. Soc. 402, 381–385, 2010, doi:10.1111/j.1365-2966.2009.15883.x
15. Gahm, G. et al., "Globulettes as seeds of brown dwarfs and free-floating planetary-mass objects," *AJ* **133**, 1795-1809, 2007.
16. Schmitt-Kopplin, P. et al., "High molecular diversity of extraterrestrial organic matter in Murchison meteorite revealed 40 years after its fall," *PNAS* **107(7)**, 2763–2768, 2010, doi:10.1073/pnas.0912157107.
17. Wickramasinghe, C., "The astrobiological case for our cosmic ancestry", *Int. J. Astrobiol.*, Volume **9**, Issue 02, 119-129, 2010, doi:10.1017/S14735504099990413.
18. Louis, G. & Kumar, S., "New biology of red rain extremophiles prove cometary panspermia," http://arxiv.org/abs/astro-ph/0312639.
19. Bassez, M. P., "Is high-pressure water the cradle of life?," *J. Phys.: Condens. Matter* **15,** L353–L361, 2003.
20. Gangappa, R, Wickramasinghe, N. C., Kumar, S. & Louis, G., "Growth and replication of red rain cells at 121 $^{o}$C and their red fluorescence", *SPIE* **7819** *proceedings*, 2010.
21. Moutou et al., "New PAH mode at 16.4 m", *A&A* **354**, L17, 2000.
22. Wickramasinghe, N. C., Wallis, J., Gibson, C. H., and Schild, R. E., "Evolution of primordial planets in relation to the cosmological origin of life," SPIE 7819 proceedings, 2010.
23. McCafferty, P., "Bloody rain again! Red rain and meteors in history and myth," *International Journal of Astrobiology* **7**, 2008, 9-15, doi:10.1017/S1473550407003904.